\documentclass[final, 5p,times,twocolumn,sort&compress]{elsarticle}	

\makeatletter
\def\ps@pprintTitle{%
 \let\@oddhead\@empty
 \let\@evenhead\@empty
 \def\@oddfoot{}%
 \let\@evenfoot\@oddfoot}
\makeatother

\usepackage{hyperref}
\usepackage{lineno}
\modulolinenumbers[5]

\usepackage{multirow}
\usepackage{soul}  
\usepackage{empheq} 
\usepackage{threeparttable, booktabs}
\usepackage{changepage}	 
\usepackage{color,colortbl} 		
\usepackage{amsmath, amssymb, amsfonts, bm, textcomp}

\usepackage{float, placeins, tikz, graphicx}
\usepackage{multirow, array}
\usepackage{iftex}
\usepackage[version=4]{mhchem}
\usepackage[load=abbr]{siunitx}
\DeclareSIUnit\Molar{\textsc{m}}

\newcommand\Tstrut{\rule{0pt}{2.6ex}}       
\ifXeTeX%
\else%
\fi%

\usepackage{romannum}
\definecolor{LightCyan}{rgb}{0.88,1,1}
\definecolor{light-gray}{gray}{0.9}
\usepackage{xr}
\newcommand*\sfref[1]{\ref*{#1}}
\newcommand\stref[1]{\ref*{#1}}

\usepackage[figurename=Fig.]{caption}

\newcommand{\GDME}{${g}_{\text{\ce{Li+}-DME}}(r)$}
\newcommand{\GDOL}{${g}_{\text{\ce{Li+}-DOL}}(r)$}

\newcommand{\GTFSIDME}{${g}_{\text{\ce{TFSI-}-DME}}(r)$}
\newcommand{\GTFSIDOL}{${g}_{\text{\ce{TFSI-}-DOL}}(r)$}

\newcommand{\trm}[1]{{\textrm{#1}}}

\newcolumntype{K}[1]{>{\centering\arraybackslash}p{#1}}
\DeclareSIUnit\Molar{\textsc{M}}

\begin{document}

\begin{frontmatter}

\title{Molecular simulations of electrolyte structure and dynamics in lithium--sulfur battery solvents}
\author[mymainaddress,mysecondaryaddress]{Chanbum Park}

\author[mymainaddress]{Matej Kandu\v{c}}

\author[mymainaddress,mysecondaryaddress]{Richard Chudoba}

\author[mymainaddress,mysecondaryaddress]{Arne Ronneburg}

\author[mymainaddress]{Sebastian Risse}

\author[mymainaddress,mysecondaryaddress]{Matthias Ballauff}

\author[mymainaddress,mysecondaryaddress]{Joachim Dzubiella\corref{mycorrespondingauthor}}
\cortext[mycorrespondingauthor]{Corresponding author}
\ead{joachim.dzubiella@helmholtz-berlin.de}

\address[mymainaddress]{Institut f\"ur Weiche Materie und Funktionale Materialien, Helmholtz-Zentrum Berlin f\"ur Materialien und Energie, Hahn-Meitner-Platz 1, 14109 Berlin, Germany}
\address[mysecondaryaddress]{Institut f\"ur Physik, Humboldt-Universit\"at zu Berlin, Newtonstr. 15, 12489 Berlin, Germany}

\begin{abstract}

The performance of modern lithium-sulfur (Li/S) battery systems critically depends on the electrolyte and solvent compositions. 
For fundamental molecular insights and rational guidance of experimental developments, efficient and 
sufficiently accurate molecular simulations are thus in urgent need. Here, we construct a molecular dynamics (MD) computer simulation model 
of representative state-of-the art electrolyte--solvent systems for Li/S batteries constituted by
lithium-bis(trifluoromethane)sulfonimide (\ce{LiTFSI}) and \ce{LiNO3} electrolytes in mixtures 
of the organic solvents 1,2-dimethoxyethane (DME) and 1,3-dioxolane~(DOL). We benchmark and verify our simulations by comparing structural and dynamic features with various available experimental reference systems and demonstrate their applicability for a wide range of electrolyte--solvent compositions. For the state-of-the-art battery solvent, we finally calculate and discuss the detailed composition of the first lithium solvation shell, the temperature dependence of lithium diffusion, as well as the electrolyte conductivities and lithium transference numbers.  Our model will serve as a basis for efficient future predictions of electrolyte structure and transport in complex  electrode confinements for the optimization of modern Li/S batteries (and related devices). 

\end{abstract}

\begin{keyword}
Battery \sep Electrolyte \sep Molecular Dynamics  \sep Conductivity \sep Solvation \sep Ion pairing.
\end{keyword}

\end{frontmatter}


\section{Introduction}

Lithium--sulfur (Li/S) batteries are discussed as a cost efficient key technology for future applications in portable electronic devices, electromobility, and as a backup storage system for the reliable use of renewable energies \cite{aurbach, Schipper2016, cuisinier, lindar1, lindar2, Gao, Scheers, Shin,Yin2013, Vijayakumar2014}. Because of their high theoretical electrochemical capacity of \SI{1675}{\milli\ampere\hour\per\gram}, Li/S batteries represent in principle an efficient energy storage system. Moreover, the abundance and low-cost of their raw materials are important advantages of this battery concept.  

The actual performance delivered by Li/S batteries is proving to be severely limited in many cases, which is directly related 
to the role of the electrolyte~\cite{barghamadi,Gao,Scheers,Shin,Vijayakumar2014,Li2016,xu,fujun}.
Ultimately, the successful development of the Li/S battery requires careful coordination of the choice 
of electrolyte with the specific nature of the cathode material. In particular, the optimal electrolyte 
has to fulfill several boundary conditions, as such to maximize charge carrier conductivity and high ionic dissociation but also to guarantee \ce{Li+} dissolution
   and stabilization of the lithium anode~\cite{Suo}. For the latter, the most prominent example is lithium nitrate (\ce{LiNO3}), which has been introduced to stabilize the anode by a protective layer formed on the electrode surface~\cite{mikhaylik2014electrolytes,aurbach2009surface}. 
 Recent developments have empirically demonstrated that lithium TFSI (bis(trifluoromethane)sulfonimide) 
  salts (at about 1~M concentration) in 1:1 mixtures of the organic solvents 1,2-dimethoxyethane (DME) and 1,3-dioxolane (DOL)
  are found to  be a suitable electrolyte solution for Li/S batteries, satisfying many of the requirements~\cite{Scheers, Risse:PCCP2016}.
More generally, these and similar electrolyte/solvent compositions are also relevant for the development of lithium--oxygen and lithium--air batteries~\cite{Geng, Balaish, Li2016, fujun} as well as for \ce{Na}/\ce{S} batteries~\cite{NaS}. 

For fundamental structural insights on a molecular level and rational guidance of experimental developments, efficient and 
accurate molecular simulations are of significant importance. For this, classical polarizable or non-polarizable all-atom force field simulations promise the best compromise between accuracy and efficiency \cite{Smith1993, Smith1998, Soetens, gaff, trappe, oplsaaref, 3anderson, Zhe, 16lesch, Borodin2006, seo2014solvate, borodin2006development, borodin2007li+, barbosa2017development}. For example, they can demonstrate how the details of local solvation structures or ion pairing affinities can be linked to transport properties, such as diffusion and conductivity, i.e., they establish {\it structure-property-function relationships}. In particular, they elucidate the effects of organic solvents on the lithium ion solvation and transport in ionic liquid electrolytes \cite{Zhe},  i.e., the solvate structures of \ce{LiTFSI} electrolytes~\cite{Borodin2006, seo2014solvate} as well as  the influence of cations on lithium ion coordination and transport~\cite{16lesch}. However, despite their importance for modern Li/S battery development, the simulation studies of the structural properties of the lithium salts in mixtures of DME/DOL solvents are scant~\cite{rajput2017elucidating}. Of particular interest is, for instance, an accurate structural characterization of the local lithium solvation structure, which is decisive for ion permeation and capacitance build-up within the commonly used porous organic electrode materials. 

Here, we construct a molecular dynamics (MD) computer simulation model  of representative state-of-the art electrolyte--solvent systems~\cite{Scheers, Risse:PCCP2016} for Li/S batteries constituted by \ce{LiTFSI} and \ce{LiNO3} electrolytes  in mixtures of DME and DOL. We focus on a force field without explicit many-body polarizability as often used before~\cite{borodin2006development} in order to enhance computing speed and invoke less parameters, aiming at comparable accuracy of structural and dynamic properties. We benchmark and verify our simulations by comparing those calculated structural and dynamic features with various  available reference systems, i.e., structure, density, dielectric constant, and viscosity of the organic solvents and their mixtures \cite{3anderson, 5farber, 14gurung, Giner}, as well as ionic diffusion and conductivity in dilute electrolytes in the respective pure (one-component) solvents~\cite{10hayamizu}. With these we demonstrate their applicability for a wide range of electrolyte--solvent compositions.  As a first practical demonstration at hand of the state-of-the-art electrolyte solvent for Li/S batteries, we calculate and discuss the detailed composition of the first lithium solvation shell, the temperature dependence of lithium diffusion, as well as the electrolyte conductivities and lithium transference numbers.  We complement this with new experimental measurements on viscosity and conductivity. Our validated model will thus serve as a basis for efficient future predictions of electrolyte structure and transport in complex  electrode confinements for the optimization of modern Li/S batteries (and related devices). 

This paper is organized as follows: we first describe in detail the used simulation model and underlying analyzing numerical methods for evaluations of relevant quantities. In Section~\ref{sec:experiment} we describe experimental settings. We then present and discuss the results in Section~\ref{sec:results-main}, followed by concluding remarks in Section~\ref{sec:conclusions}.

\begin{figure}[h]
\begin{center}
\includegraphics[width=0.9\linewidth]{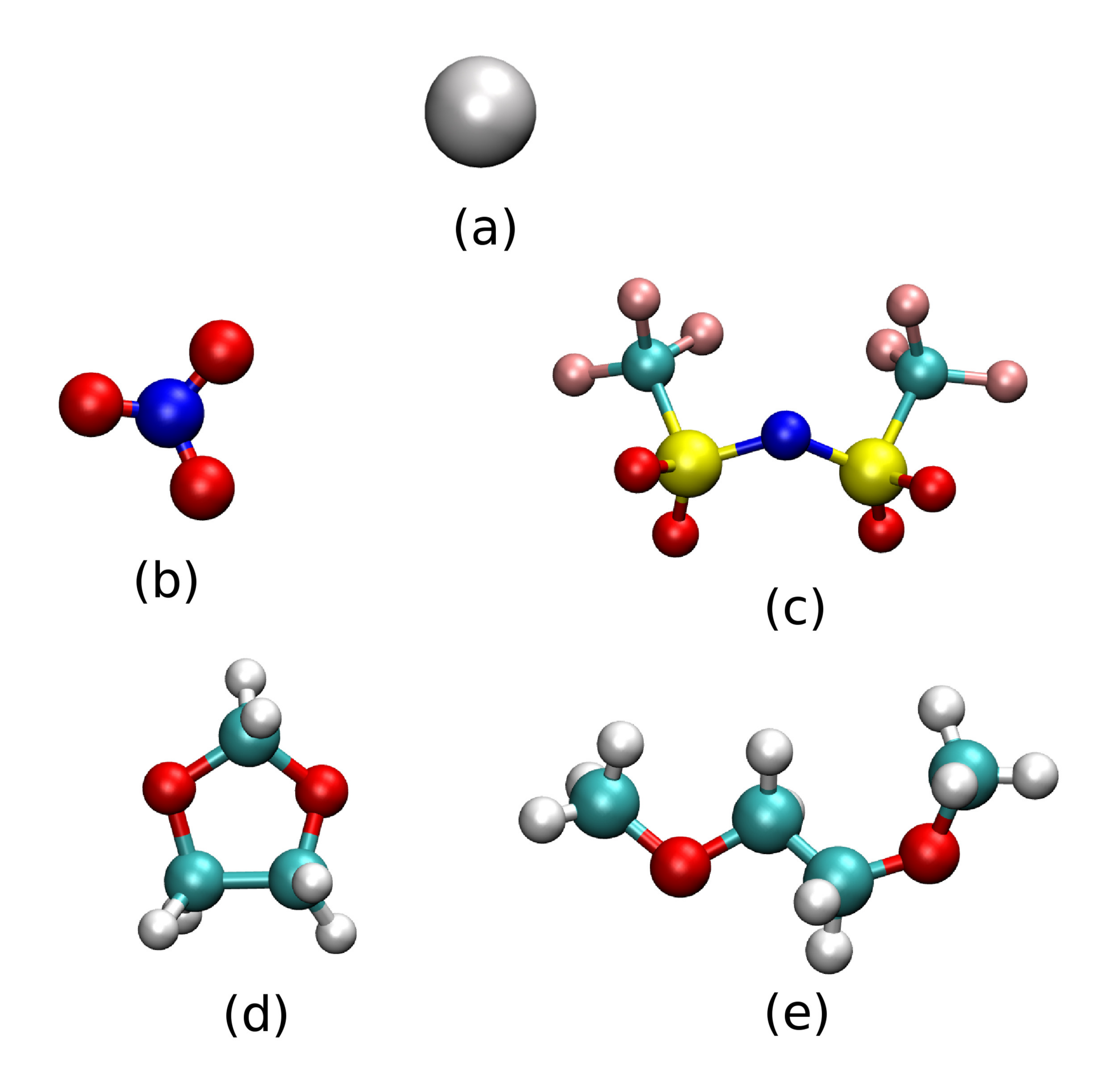}
\caption{Exemplary illustrations of the ions and solvent molecules involved in our simulation study: Ions: (a)~\ce{Li+}, (b)~\ce{NO3-}, (c)~\ce{TFSI-} and organic solvent molecules: (d)~DOL and (e)~DME. The sizes of \ce{Li+} and \ce{NO3-
} are scaled up for aesthetic reason.
}
\label{fig:ss-molecules}
\end{center}
\end{figure}
\section{Computer simulations and analysis methods \label{sec:sec-method}}
\subsection{Simulation details}\label{sec:simulationdetails}

We perform all-atom MD computer simulations of bulk electrolytes in mixed organic solvents constituted of the molecules displayed in Fig.~\ref{fig:ss-molecules} employing the GROMACS~5.1 simulation package~\cite{gromacs-5-1}. 
The production simulations are performed in the $NpT$ ensemble at constant pressure and constant temperature in a cubic box with periodic boundary conditions in all three Cartesian directions. The temperature is maintained by the Berendsen thermostat at 304~K in system \Romannum{3} (defined further below) and 298~K for all other systems with a time constant of \SI{0.1}{\ps}~\cite{berendsen}.  A constant pressure of \SI{1}{\bar} is controlled by the Parrinello--Rahman barostat~\cite{parrinello1, parrinello2} with a coupling constant of \SI{2}{\ps}.  
Electrostatic interactions are treated using the Particle-Mesh-Ewald (PME) method~\cite{pme1, pme2} with the Fourier spacing of \SI{0.12}{\nm} and \SI{1}{\nm} real-space cut-off.
For non-neutral systems, we apply uniform neutralizing
background charge in the PME.
The molecules are represented by non-polarizable models, i.e., explicit electronic polarization effects are neglected. We consider the polarization implicitly as discussed in the force-field subsection below.
All non-bonded non-electrostatic interatomic interactions are  based on the Lennard-Jones (LJ) potential with a cut-off at \SI{1}{\nm} and shifted to zero there, together with the geometric combination rules~\cite{oplsaaref} $ \sigma_{ij} = (\sigma_{ii} \sigma_{jj})^{1/2}$ and $\epsilon_{ij} = (\epsilon_{ii} \epsilon_{jj})^{1/2}$ for the LJ size and energy parameters, respectively.
The LINCS algorithm~\cite{lincs1, lincs2} is employed for all bond constraints. The integration time step is \SI{2}{\fs}.
In order to facilitate the equilibriation of solvent conformers, we first perform simulated annealing approach, where we heat the systems to 440~K or 500~K and then cool them down to 298~K on a time interval of 2~ns in the $NVT$ ensemble.

\subsection{Force fields}\label{sec:forcefields}
First we have scrutinized the properties of pure solvents of DME and DOL using various force fields and compared them with experimental benchmarks in terms of density, dielectric constant, and viscosity. The properties of DOL were assessed with three different force fields: AMBER~\cite{cornell1995second}, the Transferable Potentials for Phase Equilibria united-atom force field (TraPPE)~\cite{trappe}, and the Optimized Parameters for Liquid Simulations all-atom force-field (OPLS-AA)~\cite{oplsaaref}.  For DME, only AMBER and OPLS-AA force fields were applicable. For DME, instead of the standard dihedrals in the latter two force-fields, we implemented the optimized dihedral parameters as suggested by Anderson {\it et al.}~\cite{3anderson} based on the comparison with experimental measurements of the molecular conformation populations.  The torsional degrees of freedom lead to the occurrence of many different equilibrium conformers and thus significantly affect the instantaneous molecular multipole.  The reproduction of the correct mean static dielectric constant of DME solutions is therefore very challenging for computer simulations.  We have found that OPLS-AA (with the improved dihedrals for DME) shows consistently the best overall performance for both solvents as discussed below in the Results section.  In the Supplementary Information (SI) in Table~\stref{ext1-table:test-ff} we summarize the MD simulations results for the solvents from all tested force fields.

For low dielectric solvents DOL and DME, extra care has to be taken how to include electronic polarizability effects, which account already for more than 20\% of the total static permittivity and thus cannot be neglected. 
In view of large scale future applications, such as solvent in electrode confinement, we want to avoid introducing additional parameters into the model as well as increased intensity of the simulations. We therefore opt for an implicit inclusion of electronic (high-frequency) polarizability contribution by the Molecular Dynamics Electronic Continuum (MDEC) model~\cite{1igor, 11igor, 2kann}. It includes the electronic polarizability implicitly by  replacing all partial charges $q_i$ of ions in the simulations 
by effective, rescaled, charges $q^\trm{eff}_i$, according to 
\begin{equation}\label{qscaled}
q_i^\trm{eff} = \frac{q_i}{\sqrt{\epsilon_{\infty}}} \,.
\end{equation}
Here, $\epsilon_{\infty}$ is the high-frequency contribution to the solvent permittivity stemming from electronic fluctuations in the solvent molecules~\cite{sol}. It can be related to the refractive index $n$ as $\epsilon_{\infty}=n^2.$
From the refractive indices $n$ of 1.3781 for DME and 1.3992 for DOL~\cite{sol}, we obtain
 an effective charge of a monovalent ion of 0.73 in DME and 0.71 in DOL using eq.~(\ref{qscaled}).  We further assume that the effective charge in a mixture of DME and DOL is given simply via a linear interpolation between the effective charges in the pure DME and the pure DOL solutions. We apply eq.~(\ref{qscaled}) to all the partial charges of the three considered ions in our study, \ce{Li+}, \ce{NO3-}, and \ce{TFSI-}. The force-fields parameters of \ce{TFSI-} and \ce{NO3-} are taken from Refs.~\cite{ff-tfsi, ff-no3}.  
On the case of \ce{Li+}, we have tested various established LJ parameters and compared them with experimentally available diffusion coefficients of dilute \ce{LiTFSI} electrolyte in pure DME and DOL solvents (later defined as system III)~\cite{10hayamizu}.
While we found that the spread among the performance of the various force fields for the lithium ion is small, i.e., within 15\% for the diffusion coefficient, the best overall performance in combination with the opted anionic force-fields was exhibited by the lithium force field by Dang {\it et al.}~\cite{Dang-Li}. Hence the latter has been finally chosen for all our investigated systems with rescaled charges as defined above (see Tables~\stref{ext1-table:LJpars} and \stref{ext1-table:S-d-litfsidiox-q-opls}).  

\subsection{Simulated systems}

We simulate eight different solution `systems', denoted in the following as systems I, IIa, IIb, IIc, IIIa, IIIb, IVa, and IVb with the particular number of ions and molecules in the simulation box summarized in Table~\ref{table:moles}. System \Romannum{1} does not include ions and consists only of a reference binary mixture of DME and DOL of varying composition.  We express the organic solvent composition as the molar fraction $x$ of DOL in the solvent,
\begin{equation}
x = \frac{N_\textrm{DOL}}{N_\textrm{DOL}+N_\textrm{DME}},
\end{equation}
where $N_\textrm{DOL}$ and $N_\textrm{DME}$ correspond to the number of DOL and DME molecules, respectively. In the system class II we additionally include one \ce{Li+} (IIa), or one \ce{Li+}-\ce{NO3-} pair (IIb), or one \ce{Li+}-\ce{TFSI-} pair (IIc) to investigate diffusion and structural properties in the high dilution limit of electrolyte, also for various ratios $x=0$ to 1.  Systems \Romannum{3}a and \Romannum{3}b relate to an experimental study where diffusion coefficients and conductivity were accurately measured~\cite{10hayamizu} and consists of 25 \ce{Li+}-\ce{TFSI-} ion pairs in either  500 DME or 500 DOL solvent molecules, respectively. The molar ratio between salt and solvent is thus 1:20 in this system.  Finally, systems \Romannum{4}a and IVb represent experimental state-of-the-art compositions for a few modern batteries under development and consider concentrated electrolyte mixtures of \ce{Li+}, \ce{NO3-}, and \ce{TFSI-} at molar concentrations, given in Table~\ref{table:moles}. System IVb has a similar ionic strength as system IVa but contains no nitrate ions. 

\begin{center}
 \begin{table}[h]
 \centering
\caption{Numbers of ions and solvent molecules (i.e., the composition) in the investigated simulation systems. System~I does not contain ions and the molar ratio $x=N_{\rm DOL}/(N_{\rm DOL}+N_{\rm DME})$ of DME/DOL is varied between 0 and 1 with a total number of 508 solvent molecules. System class II has only one Li$^+$ ion (IIa) or one ion pair (IIb and IIc) for various DME/DOL ratios. System class III features a 1:20 \ce{LiTFSI} molar ratio in either DME (IIIa) or DOL (IIIb). Systems IVa and b are representative experimental state-of-the-art systems~\cite{Scheers, Risse:PCCP2016} for Li/S batteries with a molar composition of (a) \SI{0.66}{\Molar} \ce{LiNO3}, \SI{0.33}{\Molar} \ce{LiTFSI}, \SI{4.94}{\Molar} DME, and \SI{6.03}{\Molar} DOL~\cite{risse2016multidimensional} and (b) \SI{0.88}{\Molar} \ce{LiTFSI}, \SI{4.64}{\Molar} DME, and \SI{5.67}{\Molar} DOL~\cite{Safari}. 
}
  \begin{tabular}
 {@{}| c | c | c | c | c |@{}}
\hline
System   & \ce{Li+} & \ce{NO3-} & \ce{TFSI-} & solvent (DME\; /\; DOL)\\
 \hline
  \Romannum{1}&	   &	&	& 0..508\;/\;0..508\\ \hline
  \Romannum{2}a	   &1	&	&	& 0..508\;/\;0..508\\\hline
  \Romannum{2}b	   &1	&1	&	& 0..508\;/\;0..508\\\hline
  \Romannum{2}c	   &1	&	&1	& 0..508\;/\;0..508\\\hline
  \Romannum{3}a	   &25	&	&25	& 500\;/\;0\\\hline
  \Romannum{3}b	   &25	&	&25	& 0\;/\;500\\\hline
  \Romannum{4}a    &90	&60	&30	& 450/550\\ \hline
  \Romannum{4}b    &85	&	&85	& 450/550\\
\hline
  \end{tabular}
\label{table:moles}
\end{table}
\end{center}

\subsection{Analysis}\label{sec:analysis}
\subsubsection{Dielectric constant}
The non-electronic part of the static dielectric constant in the simulations is computed based on fluctuations of the Neumann dipole moment $\bf M$ of the system, 
which is the sum of all molecular dipole moments ${\bf M} = \Sigma_{i} \boldsymbol{\mu}_i$~\cite{dielectric1, dielectric2, dielectric3}. Since we are employing the MDEC treatment in our simulations, where the ionic charges are rescaled via eq.~(\ref{qscaled}), the total dipole is ${\bf M} = {\bf M}_\trm{MD} \sqrt{\epsilon_{\infty}}$, where ${\bf M}_\trm{MD}$ is the dipole moment obtained from the simulations.
The dielectric constant $\epsilon_{\mathrm{MD}}$ is calculated via~\cite{neumann1984consistent}
\begin{equation}\label{eq:dielectric}
   \epsilon_{\mathrm{MD}} = 1 + \frac{4 \pi}{3Vk_\mathrm{B}T} \bigl( \langle {\bf M}_\trm{MD}^2 \rangle - \langle {\bf M}_\trm{MD}\rangle^2\bigr),
\end{equation}
where $V$, $k_{\mathrm{B}}$, $T$ are the volume of the simulation box, the Boltzmann constant, and the absolute temperature, respectively. The dielectric constant can be affected by the size of the simulation box~\cite{saiz2000dielectric}.
We have verified that the smallest simulation box size of our system is large enough,
such that the dielectric constant is not affected by the finite-size effects (see Fig.~\sfref{ext1-fig:epsilon_md}).
Finally, we calculate the total static dielectric constant $\epsilon$ for the comparison to experimental reality as~\cite{Leontyev2011, 11igor, dielectric1, 2kann}
\begin{equation}\label{eq:epsilon}
 \epsilon = \epsilon_{\infty}  \epsilon_{\mathrm{MD}}.
\end{equation}

\subsubsection{Viscosity}
\label{sec:viscosity}
The shear viscosity $\eta_\infty$ is calculated from transverse current correlation function~\cite{Palmer,GMX, Hess}, using the transverse momentum fields (transverse-current autocorrelation function) as implemented in the GROMACS simulation package~\cite{gromacs-5-1}.
Here, a total 16 transverse-current autocorrelation functions corresponding to different $k$-vectors are
considered, resulting in 16 values of $\eta$. The values of $\eta$ are fitted to $\eta(k) =\eta_{\infty}(1 - ak^2)$, yielding the shear viscosity $\eta_{\infty}$.

In cases of binary solvent mixtures, as we investigate here, we compare the results for the viscosity to the values obtained via well-established semi-empirical analytical mixing rules by Fort and Moore~\cite{fort1966} for the experimentally expected viscosity. There, the viscosity $\eta_\trm{mix}$ of a mixture is calculated from the viscosities of the pure components
$\eta_{\infty, 1}$ and $\eta_{\infty, 2}$ as
\begin{equation}\label{eq:mixing}
\eta_{\mathrm{mix}} (\phi_1,\phi_2) = \eta_{\infty, 1}^{\phi_1} \eta_{\infty, 2}^{\phi_2}.
\end{equation}
The volume fractions $\phi_1$ and $\phi_2$ of each of the components, $\phi_i=V_i/(V_1+V_2)$, are obtained from the partial volumes $V_i$ calculated as $V_i=m_i/\rho^0_i$ with $\rho^0_i$ being the mass density of the pure component.

\subsubsection{Long-time self-diffusion coefficient}
The long-time self-diffusion coefficients of the molecules in the simulations are calculated from the mean square displacement relation,
\begin{equation}\label{eq:diffusion}
 D_\trm{MD} = \lim_{\Delta t \to \infty} \frac{ \big\langle r^2(\Delta t) \big\rangle}{6 \Delta t}.
\end{equation}
Due to finite size effects of a simulation box, hydrodynamic corrections can play a significant role.
It has been shown that hydrodynamic perturbations in a small periodic box lead to a finite size correction (FSC) of the diffusion coefficient, 
$\Delta D_{\rm FSC}$.
In the leading order, $\Delta D_{\rm FSC}$ scales inversely with the length $L$ of the simulation box and the viscosity $\eta$ of the solution, 
 $\Delta D_{\rm FSC} \propto 1/(\eta_{\infty} L)$ ~\cite{15inchul}.
Although the scaling prefactor of the latter expression can be analytically calculated based on the hydrodynamic self-interaction of a point perturbation, deviations occur for larger and more complex solutes~\cite{yeh2004diffusion}.
Therefore, we estimate the correction in our system by simulating the diffusion in boxes of various sizes and fitting its dependence to $\Delta D_{\rm FSC} =c/ L$ with $c$ as a fitting parameter (Figs.~\sfref{ext1-fig:Dfsc-sysI}, \sfref{ext1-fig:Dfsc-sysIII}, and \sfref{ext1-fig:Dfsc-sysIVa}~in the SI). 
We determine the coefficient $c$ only for the diffusion of the solvent components, DME and DOL, as there the statistical accuracy is highest, and apply the correction also to the diffusion coefficients of ions. The diffusion coefficient we compare to experiments is finally given by 
\begin{equation}
D = D_{\mathrm{MD}} +  \Delta D_{\mathrm{FSC}}.
\end{equation}

\subsubsection{Conductivity and transference number}

The stationary linear response conductivity $\sigma$ is defined by Ohm's law 
\begin{equation}
J = \sigma E, 
\label{ohm}
\end{equation}
(expressed here simply as scalar quantities), where $E$ is the external electrostatic driving field and $J=\sum_i J_i  = \sum_i \sigma_i E$ the sum of all ionic current densities, which define individual partial ionic conductivities $\sigma_i$.  The transference number of an ion $i$  is defined by~\cite{transference}
\begin{equation}
 t_i = \frac{J_i}{J} = \frac{\sigma_i}{\sigma}, 
\end{equation}
and describes the relative contribution of the current of species $i$ to the total current.
By applying an external field $E$ in the range between \SIrange[range-phrase={ and },range-units=single]{0.005}{0.05}{\volt\per\nm} (for which we verified to lie within the linear response regime, see SI~Figs.~\sfref{ext1-fig:J-E-III} and \sfref{ext1-fig:J-E-IVab})
 and calculating resulting individual currents $J_i$, we obtain partial conductivities $\sigma_i$ via eq.~(\ref{ohm}).
 By measuring the mean drift velocities $v_i$ of each ionic species under the external field, we calculate the current densities as $J_i = z_i e n_ i v_i$, where $n_i$ is the ionic number density and  $z_i$ and $e$ the valency and the elementary charge, respectively.
 In the ideal ion limit one expects the Nernst--Einstein relation to be applicable. It relates the total conductivity and diffusion coefficients, $\sigma^{\rm id} = \sum_i |z_i|eD_in_i/k_\trm{B}T$, and thus gives a reference to conductivity values regarding ideal ionic transport behavior. A degree of ion uncorrelated motion can be defined by $\alpha = \sigma/\sigma^{\rm id}$. It was shown that the parameter $\alpha$ is tightly related to ion-pair formations~\cite{borodin2007li+}, whereby smaller values imply higher degree of pair formation. 

\subsubsection{Coordination number}
The coordination number of molecules of type $i$ in the first solvation shell surrounding a single molecule of type ${j}$ is calculated as 
\begin{equation}
 N_{i} = 4 \pi n_{j} \int_0^{R_\mathrm{M}} {g}_{{ij}}(r) \, r^2 \, \mathrm{d}r,
\end{equation}
where $R_\trm{M}$ is the distance of the first minimum following the first peak in the radial distribution function~(RDF), $g_{ij}(r)$, which is a standard approach for bulk liquids~\cite{HansenMcDonaldbook}.

\section{Experimental characterization of system IVa} 
\label{sec:experiment}
The gravimetric density of the samples was measured at \SI{28}{\degreeCelsius} and at a pressure of \SI{1.0186}{\bar} using a chempro/PAAR DMA 602 density meter with a Julabo F25 thermostat. 
The average value of ten measurements of the natural frequency of a glass tube filled with the solution was taken to calculate the density. Millipore water and air served as reference for this calculation. The viscosity was determined using a Capillary Viscometer (SI Analytics 50101/0a) and a laboratory stopwatch. The viscosity was averaged over three measurements that were performed in an argon filled glovebox. 
The conductivity was evaluated by performing an impedance spectroscopy in the frequency range of  100~mHz to 1~MHz with 5~mV RMS voltage signal and 15 points per decade. A GAMRY interface 1000 potentiostat and an in-house-designed electrochemical cell were used. The cell consists of aluminum-electrodes and a cell housing made of PEEK (Polyether Ether Ketone). The electrode distance is 1~mm, the circular electrode area is 198.56 mm$^2$ (15.9~mm in diameter). The electrolyte was filled in through a hole on the side of the PEEK cell housing to ensure complete filling. The conductivity was determined by the intersection of the impedance with the x-axis at high frequencies in a Nyquist-plot. 

\section{Results and discussion \label{sec:results-main}}
\subsection{System I : Pure solvent (DME/DOL) mixtures \label{sec:results-1}}

\begin{figure}[h!]
\begin{center}
\includegraphics[width=0.8\linewidth]{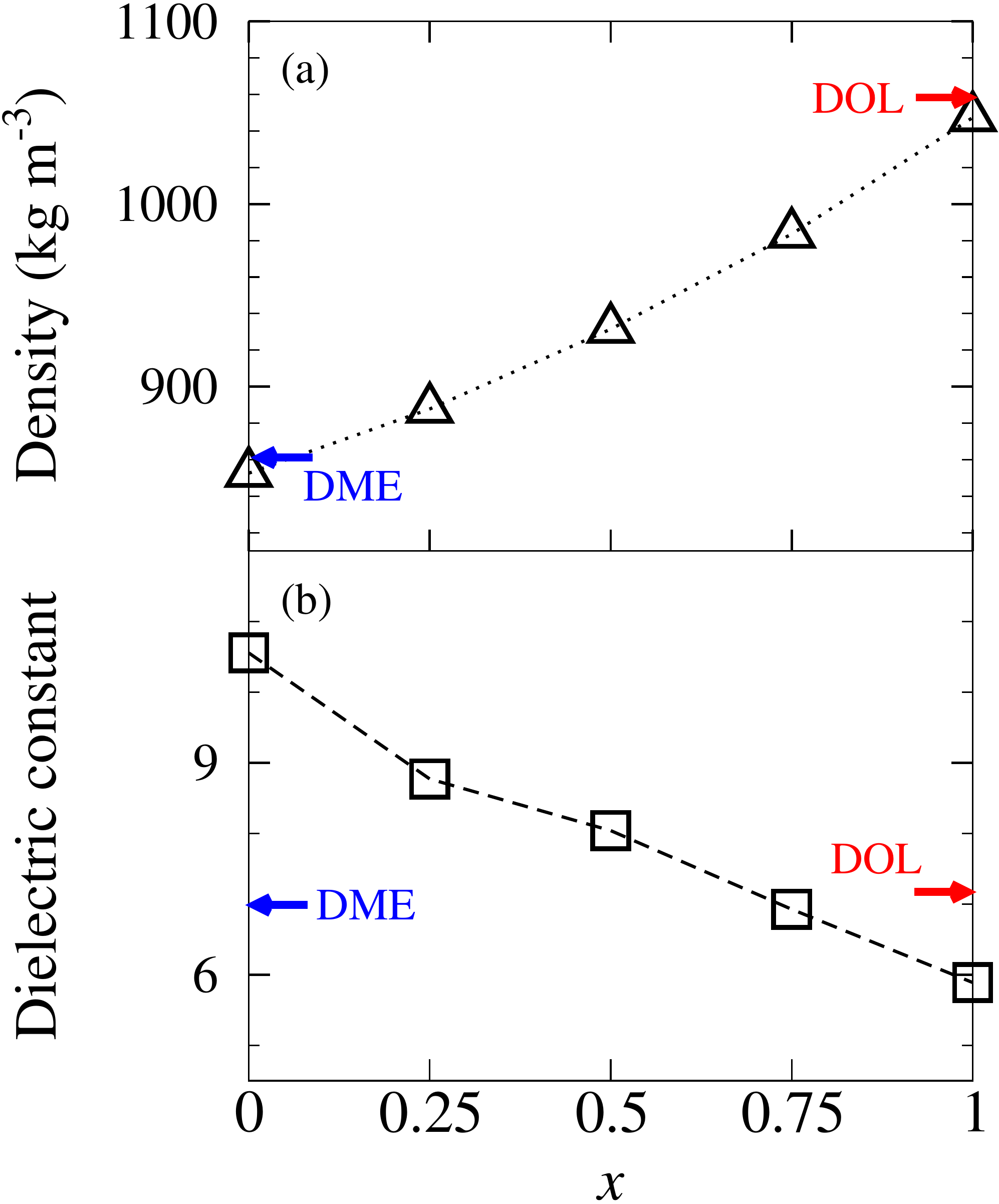}
\caption{(a) Density of the DME/DOL mixture (system \Romannum{1}) versus the molar fraction of DOL composition $x = {N_\textrm{DOL}}/({N_\textrm{DOL}+N_\textrm{DME}})$  from our MD simulations (triangular symbols).  (b) Dielectric constant  (square symbols) from the MD, eq.~(\ref{eq:epsilon}), for the same systems as in (a). The colored arrows indicate the experimental reference values of pure DME~\cite{14gurung} and pure DOL~\cite{Giner}, respectively.}
\label{fig:sys1-density}
\end{center}
\end{figure}

The density and dielectric constant of the DME/DOL mixtures as a function of the molar fraction $x$ of DOL are shown in Fig.~\ref{fig:sys1-density}(a) and (b), respectively. The density of the pure DME (i.e., $x=0$) obtained from MD is \SI{853}{\kg\per\cubic\meter}, which very well agrees with the experimental value of \SI{861}{\kg\per\cubic\meter}~\cite{14gurung}. Also the density of the pure DOL (i.e., $x=1$) from MD, \SI{1047}{\kg\per\cubic\meter}, is in good agreement with the experimental one, \SI{1059}{\kg\per\cubic\meter}~\cite{Giner}. The density of the mixture increases monotonically with the molar fraction $x$ of DOL.
Turning to the dielectric constant in panel (b), we find satisfactory agreement for the pure DOL solvent at $x=1$, where the simulated value is about 16\% smaller than in the experiments.  The MD value for the pure DME ($x=0$) is less satisfying and with 10.6 compared with the experimental 7.1 almost 50\% too large. However, after having examined and thoroughly scrutinized various force field combinations, we found this deviation still to be minimal under the constraint that the density as well as the viscosity (shown below) reproduce well the experimental reality.

\begin{figure}[h!]
\begin{center}
\includegraphics[width=0.8\linewidth]{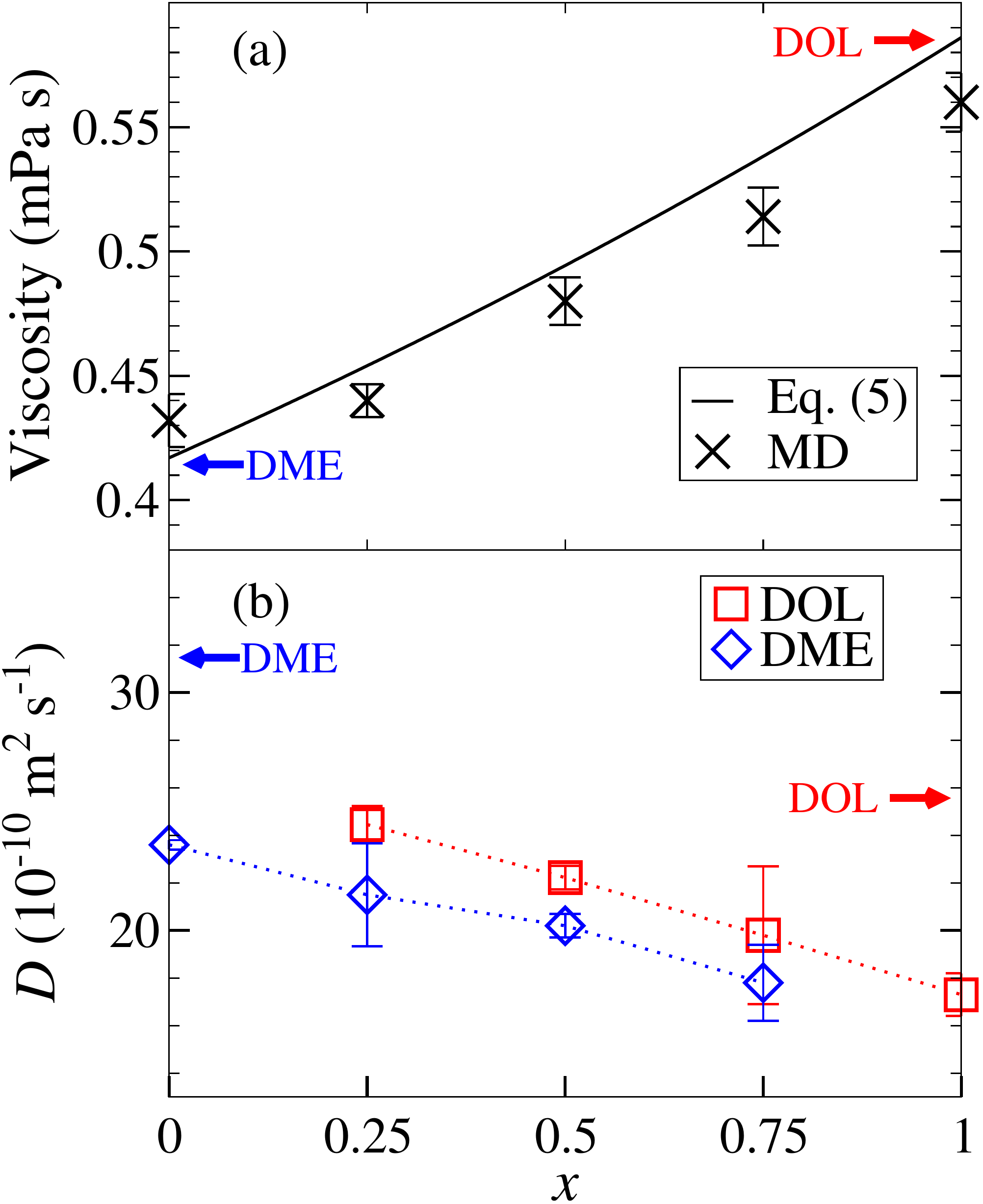}
\caption{(a) Shear viscosity $\eta(x)$ of the binary DME/DOL mixtures versus the molar fraction of DOL $x$.
Crosses indicate the results obtained from the MD simulations and the solid line represents the viscosity of the binary mixture from eq.~(\ref{eq:mixing}), which interpolates the experimental limits of the pure DME~\cite{14gurung} and pure DOL~\cite{Giner}, indicated by arrows. (b)~Self-diffusion coefficients of  DME molecules (diamonds) and DOL molecules (squares) in the DME/DOL mixture fas a function of the DOL molar fraction $x$.  The experimental reference values of DME and DOL~\cite{10hayamizu} are indicated by a blue and a red arrow, respectively.}
\label{fig:sys1-viscosity}
\end{center}
\end{figure}

The viscosity of the DME/DOL mixtures from our simulations is presented in Fig.~\ref{fig:sys1-viscosity}(a). There, we also plot experimentally measured viscosities or pure solvents and apply the analytical mixing rule eq.~(\ref{eq:mixing}) for the mixtures.
 We see that the simulations for the pure DME and DOL systems yield viscosities of \SIlist{0.43;0.56}{\milli\pascal\second}, respectively, which compare  well with the experimental ones of \SIlist{0.42;0.58}{\milli\pascal\second}~\cite{Giner, 14gurung},  as well as with the expected interpolation behavior, eq.~(5). As another important transport property we have calculated the self-diffusion coefficients of the solvent molecules in the mixtures, cf.\ panel (b) of the same figure.  The simulations, corrected for finite-size effects (cf.\ SI),  underestimate the experimental reference values by 27\% (DME) and 32\% (DOL). The interpolation between the limiting cases $x=0$ and $x=1$ transits monotonously. However, as we will see later, the diffusion coefficients compare much better to the experimental values of system \Romannum{4}b (Table~\ref{table:sys34}), performed by a different group.

In essence, we can conclude that transport properties are well captured within the MD model. Overall, we find satisfying behavior of our solvent force field in the sense that it can reproduce reasonably well several experimentally important equilibrium thermodynamic and transport properties for the full molar ratio range $x=0$ to $x=1$ at the same time. 


\subsection{Systems IIa,\,IIb,\,and IIc: single ions or ion pairs in mixed DME/DOL solvents \label{sec:results-2}}

\begin{figure}[h!]
\begin{center}
\includegraphics[width=0.8\linewidth]{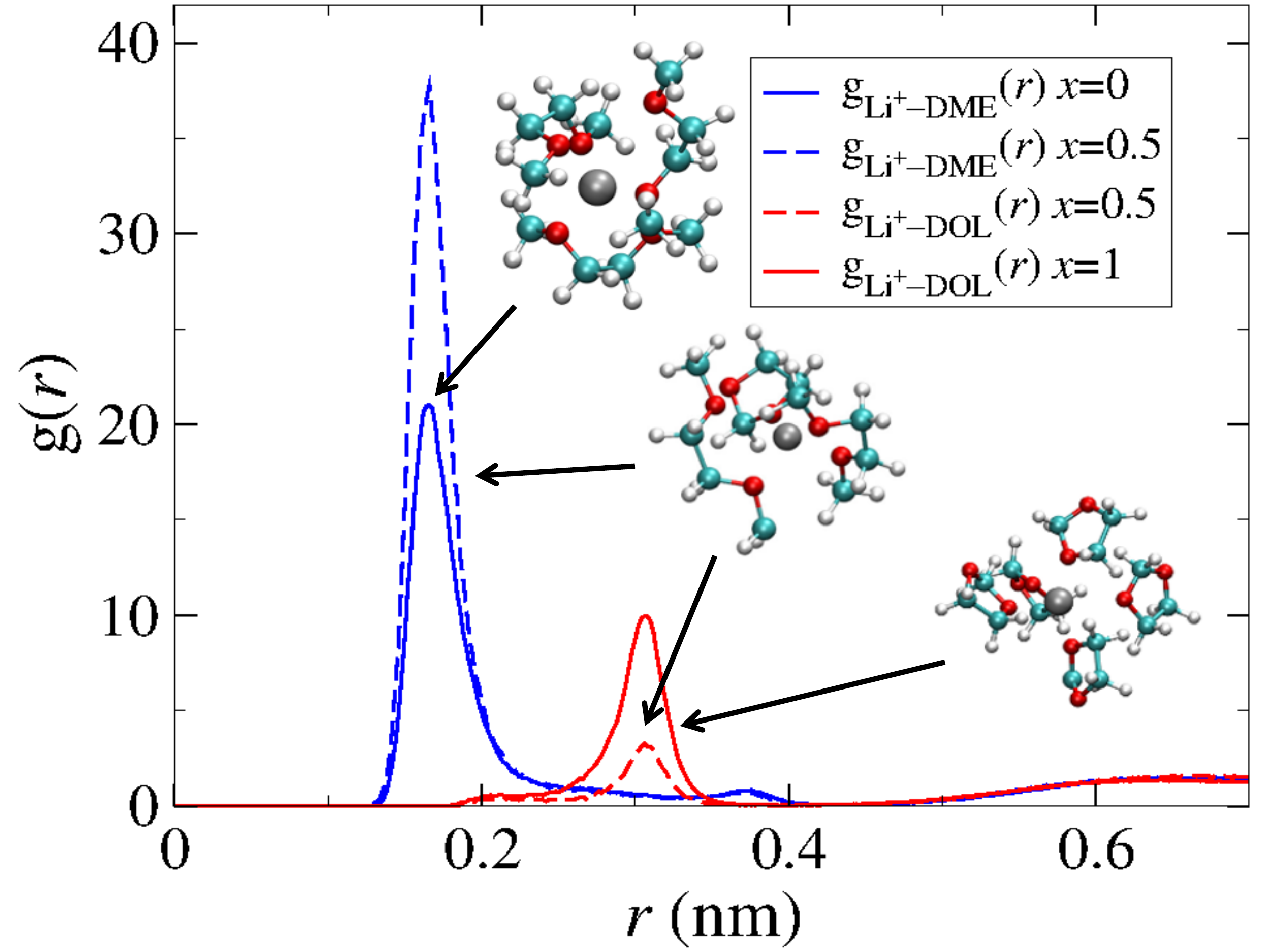}

\caption{Center-of-mass RDF between \ce{Li+} and solvent molecules for system IIa; \GDME\ and \GDOL\ are shown for the limiting cases $x=0$ (DME only) and $x=1$ (DOL only) and for the intermediate ratio $x=0.5$.
}
\label{fig:sys2a-rdf}
\end{center}
\end{figure}

Now we consider highly dilute electrolyte solutions, where we investigate 
 the solvation structure and diffusion of a single \ce{Li+} ion (Sys. IIa), \ce{Li+}-\ce{NO3-}~(IIb), and \ce{Li+}-\ce{TFSI-}~(IIc) ion pairs in the solvent mixture of system I. In order to discuss the solvation structure, we plot in Fig.~\ref{fig:sys2a-rdf} the center-of-mass radial distribution function (RDF) between the \ce{Li+} and solvent molecules in system IIa: \GDME\ and \GDOL\  are  shown for the limiting cases $x=0$ (DME only) and $x=1$ (DOL only) and for the intermediate ratio $x=0.5$. The DME distribution peaks at about 0.16~nm and is thus closer to the \ce{Li+} ion than DOL molecules, whose distribution peaks at about 0.3~nm. Such a close approach of DME is consistent with experimental data where the \ce{Li+}-DME coordination leads to {\it cis} (the C-O bonds) and a {\it gauche} configurations in DME molecules in a bidentate binding configuration~\cite{brouillette, Henderson2006glyme}. This is absent for DOL, cf.\ also the representative simulation snapshots in Fig.~\ref{fig:sys2a-rdf}.  Thus, the coordination of DME in the bidentate to \ce{Li+} ion retains a relatively stable solvation structure even at the symmetric solvent ratio $x=0.5$. 

\begin{figure}[h!]
\begin{center}
\includegraphics[width=0.8\linewidth]{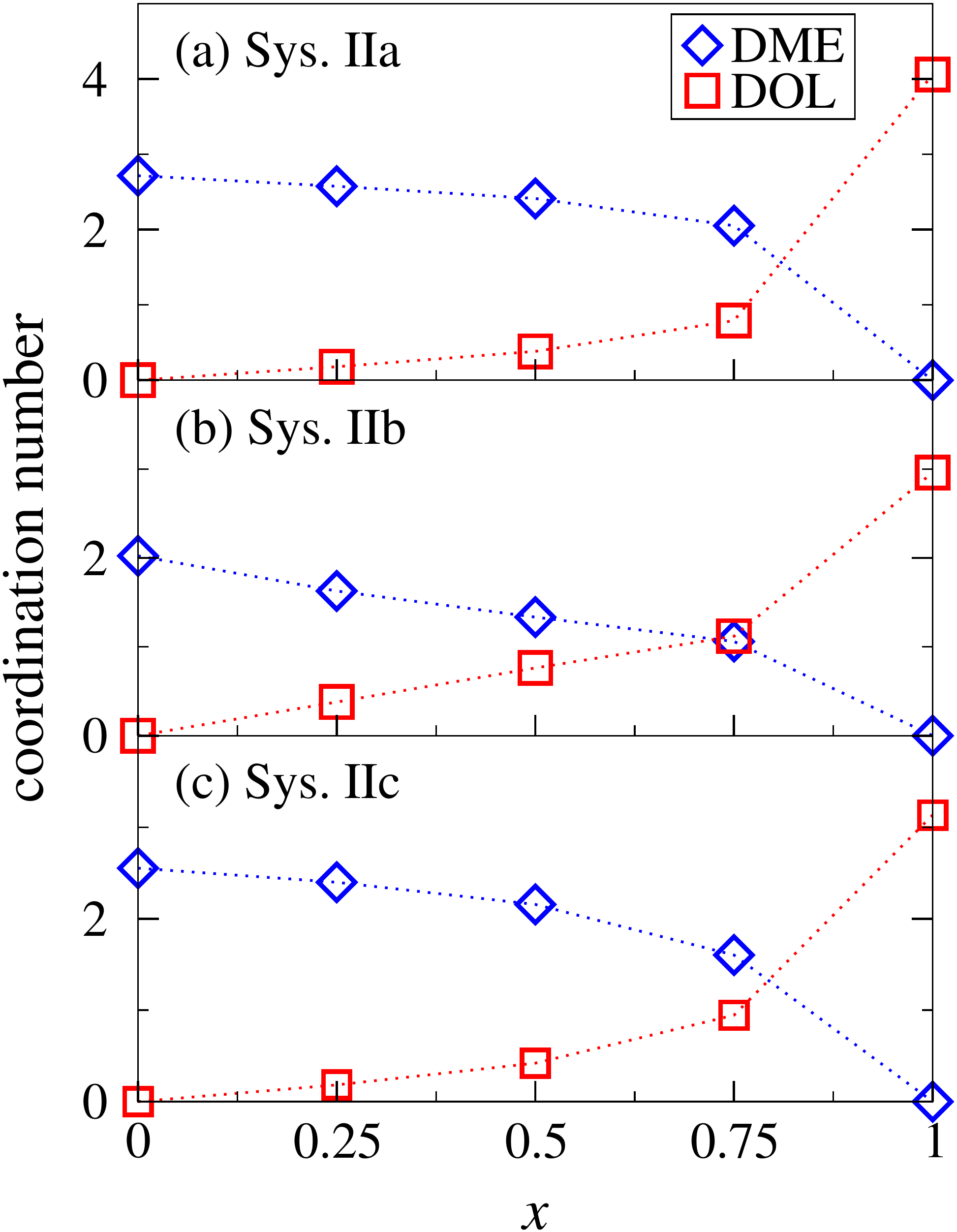}
\caption{Coordination number of \ce{Li+} of DME (blue diamonds) and DOL (red squares) molecules as a function of the molar fraction $x$ in systems IIa, IIb, and IIIc. The dotted connecting lines are plotted as guides to the eye.}
\label{fig:sys2-coordination}
\end{center}
\end{figure}
The consequence is an interesting coordination behavior along the mixing coordinate $x$ as presented in Fig.~\ref{fig:sys2-coordination}(a) for system IIa: let us start at the right hand side of the plot at $x=1$, where DOL coordinates the cation with a coordination number of 4. Adding DME to the solution very quickly substantially changes the DOL coordination; already at around $x\simeq 0.8$ the coordination of DME and DOL equalize (at about 1.8). At a symmetric concentration ($x=0.5$) the DME is then in large excess with a coordination close to the limiting coordination of about 2.7 of the pure DME ($x=0$).   We note that an analogous `solvent-exchange' behavior has been observed previously already in DME/propylene carbonate mixtures~\cite{chaban}, pointing to the special excess solvation properties of DME in general for its mixtures with other solvents. 

For systems IIb and IIc, where also an additional anion is present, corresponding to concentration around \SI{20}{\milli\Molar} of the electrolyte, the cation coordination number decreases, but qualitatively retaining the behavior with varying the molar fraction $x$, see Fig.~\ref{fig:sys2-coordination}(b) and (c).  The reason is a strong ion pairing, which is anion-specific. The relatively small \ce{NO3-} counterion binds very tightly to the \ce{Li+} cation (see also the discussion later for the concentrated system IVa). The strongly associated \ce{Li+}-\ce{NO3-} ion pairs in the pure DME indeed have been categorized previously already as a `contact ion pair' (CIP) or even `aggregate' solvation structure~\cite{Henderson2006glyme}. In this case, the coordination by the organic solvent is consequently reduced by 1 in the whole $x$-range. The larger \ce{TFSI-} anion, however, consistent with the category of a dissociated salt forming `solvent-separated ion pairs' (SSIPs) in pure DME~\cite{Henderson2006glyme}, only manages to replace bigger more weakly bound DOL molecules, but not DME molecules. Consequently, the DME coordination around \ce{Li+} close to $x=0$ remains almost unaffected by the presence of the \ce{TFSI-} ion. The SSIP structure of \ce{Li+}-\ce{TFSI-} can also be empirically explained with the Gutmann donor number~(DN)~\cite{Gutmann, gutmann1976empirical, ueno2012glyme}, 
which corresponds to the negative binding enthalpy of a given molecule to a reference Lewis acid. Assuming \ce{Li+} to behave as a Lewis acid, the DN provides an estimate for the binding affinities of other molecules to \ce{Li+}. DME has the DN number of 20, while \ce{TFSI-} the value of 5. Hence, much weaker association strength of \ce{Li+}-\ce{TFSI-}, as indicated by the smaller DN number, leads to the solvation shell of \ce{Li+} dominated by DME (with higher DN). Our results are in line with the reported ionic strength in aprotic solvents in general, where the \ce{TFSI-} has much smaller association strength than \ce{NO3-}~\cite{Henderson2006glyme}.

\begin{figure}[h!]
\begin{center}
\includegraphics[width=0.8\linewidth]{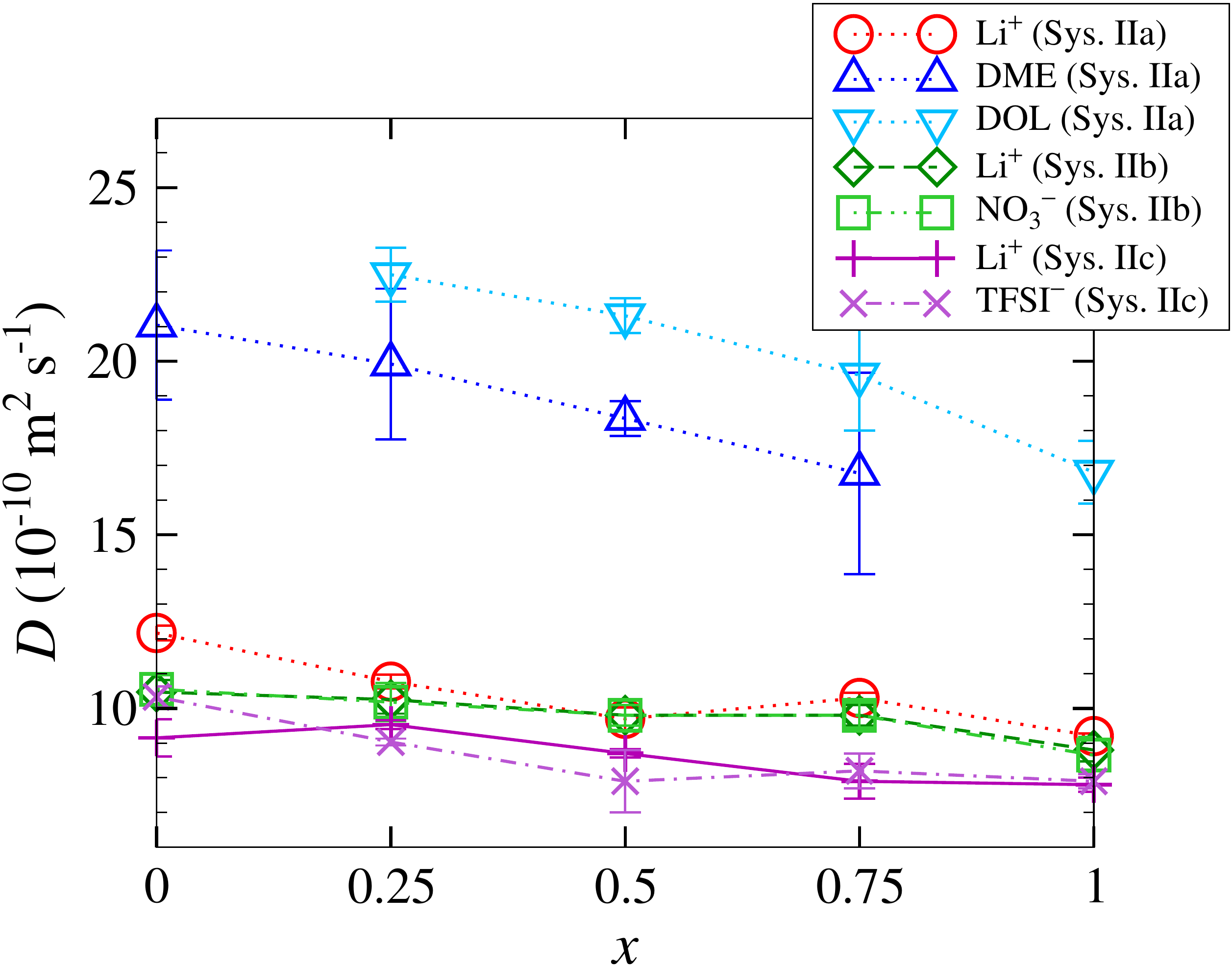}
\caption{Diffusion coefficients of \ce{Li+}, \ce{NO3-}, \ce{TFSI-}, DME, and DOL in systems \Romannum{2}a, \Romannum{2}b, and \Romannum{2}c as a function of the DOL molar fraction $x$.}
\label{fig:sys2a-d}
\end{center}
\end{figure}

 \begin{table*}[ht!]
 \begin{threeparttable}
\caption{Density, dielectric constant, viscosity, \ce{Li+} coordination, conductivity and diffusion coefficients of systems \Romannum{3}a (1:20 \ce{LiTFSI} salt in pure DME), \Romannum{3}b (1:20 \ce{LiTFSI} salt in pure DOL), \Romannum{4}a [\SI{0.66}{\Molar} \ce{LiNO3} and \SI{0.33}{\Molar} \ce{LiTFSI} in DME:DIOX (45:55 molar ratio)] and \Romannum{4}b [\SI{0.88}{\Molar} \ce{LiTFSI} in DME:DIOX (45:55 molar ratio)]. Systems \Romannum{3}a and  \Romannum{3}b are conducted at $T$\,$=$\,304~K and, whereas \Romannum{4}a and \Romannum{4}b at 298~K. 
Experimental measurements of conductivity in this work (system \Romannum{4}a) are carried out with \SI{0.6}{\Molar} \ce{LiNO3} and \SI{0.3}{\Molar} \ce{LiTFSI} in a DME and DOL (1:1 wt$\%$) mixture at room temperature.}
 \begin{tabular}
{@{}| p{4cm} || K{1.2cm} | K{1cm} || K{1.2cm}  | K{1cm} || K{1.3cm} | K{1.7cm} || K{1.5cm} | K{1cm} | @{} }


\hline
&\multicolumn{2}{c ||}{Sys. \Romannum{3}a} & \multicolumn{2}{c||}{Sys. \Romannum{3}b} & \multicolumn{2}{c||}{Sys. \Romannum{4}a} & \multicolumn{2}{c |}{Sys. \Romannum{4}b}\\
\cline{2-3}
\cline{4-5}
\cline{6-7}
\cline{8-9}
& \multirow{2}{*}{MD} & {Exp.}            & \multirow{2}{*}{MD}   & {Exp.}            & \multirow{2}{*}{MD} & {Exp.}      & \multirow{2}{*}{MD} & Exp.\\
&                     & \cite{10hayamizu} &                       & \cite{10hayamizu} &                     & (this work) &       & ~\cite{Safari, zheng2013ionic, kim2011electrochemical, rajput2017elucidating}\\
\hline\Tstrut
Density (\si{\kg\per\cubic\meter})	& 934.1(2)	& 	& 1147.0(1) 	&	& 1030.0(1) & 1103 & 1091.2(1.1) & 1125$^{a}$ \\	
Dielectric constant $\epsilon$ 		& 8.8	 	&	& 5.7	  	& 	& 7.6	    &  & 6.6 &  \\		
Viscosity (\si{\milli\pascal\second})	& 0.57(5) & & 0.74(5) &		&0.77(1) & 0.88 & 0.85(3) & 1.56$^{b}$, 1.25$^{c}$ \\	
Coordination number of \ce{Li+}         & 2.45		&	& 3.04	&	& 3.2 &  & 3.2 & \\
Conductivity  (\si{\siemens\per\meter})          & 0.73(3) & 0.89 & 0.27(3) & 0.31  & 0.36(4) & 0.59(3) & 0.82(3) & 1.47$^a$, 1.32$^{b}$, 1.1$^{c}$ \\
Degree of ion uncorrelated motion $\alpha$                     & 0.28    & 0.31 & 0.1     & 0.1   & 0.12    &        & 0.3     & \\
\ce{Li+} transference $t_{\ce{Li+}}$             & 0.4     & 0.47 & 0.5     & 0.51  & 0.61    &        & 0.38    & \\
$D_{\ce{Li+}}$ (\SI[retain-unity-mantissa=false]{1E-10}{\meter\squared\per\second}) & 7.2(1.1) & 7.7 & 5.1(4) & 6.4  	& 4.0(3) &  & 4.7(3) & 4.3$^d$\\ 
$D_{\ce{TFSI-}}$ (\SI[retain-unity-mantissa=false]{1E-10}{\meter\squared\per\second}) & 7.8(1.3) & 8.8	& 4.5(7) & 6.2  & 5.0(1.0) &   & 3.8(1.0) & 4.8$^d$\\ 
$D_{\ce{NO3-}}$ (\SI[retain-unity-mantissa=false]{1E-10}{\meter\squared\per\second})  &  & 	&  &   	& 3.9(3) &  &  & \\ 
$D_{\ce{DME}}$ \; (\SI[retain-unity-mantissa=false]{1E-10}{\meter\squared\per\second}) & 16.4(1.3) & 22.0 &  &  & 9.9(1.0) &  & 8.2(1.0) & 7.7$^d$\\ 
$D_{\ce{DOL}}$ (\SI[retain-unity-mantissa=false]{1E-10}{\meter\squared\per\second}) & & & 12.9(4) & 17.0 & 13.1(1.0) &  & 10.0(1.0) & 11.4$^d$\\ 
\hline
\end{tabular}
\label{table:sys34}
\begin{tablenotes}[flushleft]
      \footnotesize
      \item $^{a}$Ref.~\cite{Safari}, $^{b}$Ref.~\cite{zheng2013ionic}, $^{c}$Ref.~\cite{kim2011electrochemical}, and $^{d}$Ref.~\cite{rajput2017elucidating}
 \end{tablenotes}
\end{threeparttable}
\end{table*}
The self-diffusion coefficients of the molecular constituents in all systems of class II are presented in Fig.~\ref{fig:sys2a-d}. In these dilute systems (molar ratio 1:508) the diffusion properties of the pure organic solvent mixtures of DME and DOL are hardly affected. The \ce{Li+} diffusion coefficient is between \SI{13E-10}{\meter\squared\per\second} at $x=0$ and \SI{9E-10}{\meter\squared\per\second} at $x=1$ in system IIa, i.e., decreasing with increasing DOL concentration. This effect can be attributed to larger size of the solvation shell in a more coordinated DOL solvent. An inclusion of the counterion has small but visible effects and depends on anion type. For \ce{NO3-}, a strong ion pair is created, which evidently changes the \ce{Li+} diffusion only very slightly although a joint diffusion of the cation--anion pair is apparently established. In the presence of a larger \ce{TFSI-} anion, the \ce{Li+} cation diffusion is slowed down more, probably related to a larger size of the formed ion pair, although only existent as an SSIP cluster. 

\subsection{Sys. IIIa and IIIb \label{sec:results-3}: 1:20 electrolyte to solvent ratio}

Now we investigate the systems with the molar ratio of 1:20 LiTFSI:DME or LiTFSI:DOL, for which accurate experimental data for the conductivity, lithium transference, degree of ion uncorrelated motion, and diffusion coefficients are available~\cite{10hayamizu}. Other physical properties, such as density, dielectric constant, viscosity, and solvent coordination are also calculated and summarized in Table~\ref{table:sys34} together with the experimental diffusion data. As can be seen, the total conductivity, lithium transference, as well as degree of ion uncorrelated motion are quite well reproduced. The self-diffusion of DME and DOL is in the simulations lower (by about factor of 3/4) than in the experiment. This deviation is consistent with the results of the pure solvent (system I) conducted within the same experimental study~\cite{10hayamizu}. The diffusion coefficients of the \ce{Li+} and \ce{TFSI-} ions from the MD also consistently reproduce the experimental trends~\cite{10hayamizu} in both systems \Romannum{3}a and \Romannum{3}b. 
 In experiments and simulations, it is observed that the diffusivity in system \Romannum{3}a is faster than that of \Romannum{3}b.  This can be attributed to the lower viscosity of DME with respect to DOL.  Compared with the simulations results of the highly diluted electrolyte in systems II (1:500 ion--solvent ratio), the diffusivities in systems III 
are all found about 20\% (DME) to 35\% (DOL) lower  due to the higher viscosity by the same relative amount.

The coordination solvent numbers  for \ce{Li+} in systems \Romannum{3}a and \Romannum{3}b are 2.45 and 3.04, respectively (see Table~\ref{table:sys34}). The corresponding  RDFs (see Figs.~\sfref{ext1-fig:rdf-2-2}) indicate, analogously to systems II, that the  DME molecule solvates \ce{Li+} much stronger than DOL.  The distance between \ce{Li+} and the centers-of-mass of DME and \ce{Li+} and DOL are 0.16 and 0.30~nm, respectively, which is in accord with  \GDME\ and \GDOL\ for $x=0$ and $x=1$ in the molar ratio of 1:500 in Sec~\ref{sec:results-2}, respectively. \GTFSIDME ~and \GTFSIDOL ~(see Fig.~\sfref{ext1-fig:sys2c-rdf-tfsi-dmediox}) show the first peak at \SIlist{0.67;0.65}{\nm}, respectively, which are  relatively low, similarly as in the much more dilute systems~II. This implies that the affinities between \ce{TFSI-} and DME or DOL are relatively small and do not much depend on the salt concentration.

\subsection{Systems IVa and IVb: Li-ion battery electrolyte solution}

Systems IVa and IVb consider practical Li-ion battery electrolyte solutions with about \SI{0.99}{\Molar} salt concentration in a 45/55 molar ratio DME/DOL solvent. The individual molar concentrations are \SI{0.66}{\Molar} \ce{LiNO3}, \SI{0.33}{\Molar} \ce{LiTFSI}, \SI{4.94}{\Molar} DME, and \SI{6.03}{\Molar} DOL in system IVa~\cite{risse2016multidimensional} and  \SI{0.88}{\Molar} \ce{LiTFSI},  \SI{4.64}{\Molar} DME, and \SI{5.67}{\Molar} DOL in nitrate-free system IVb. 
Table~\ref{table:sys34} shows the comparison between MD and experimental results for the density and viscosity, dielectric constant, conductivity and diffusion coefficients for systems \Romannum{4}a and \Romannum{4}b. The MD simulations are able to reproduce well the available experimental values, further verifying the quality of our implemented force field. 
The MD model reproduces very well all the diffusion coefficients in system IVb (with deviations below 20\%). Moreover, in system \Romannum{4}a, it yields the conductivity of \SI{0.36}{\siemens\per\meter}, which is around 40\% lower than the experimental value of \SI{0.59}{\siemens\per\meter}, but correctly catches the trend of decrease in conductivity due to introduction of \ce{NO3-} ions (compare systems \Romannum{4}a, \Romannum{4}b; experimental conductivity measurements at different \ce{LiNO3} molar concentrations are available in Table~\stref{ext1-table:S-sigma-expt}). 

The value of the transference number of the \ce{Li+} in system IVa is about 0.6.  The relatively low value of $\alpha$ $\simeq$ 0.12, implyies that strong ion pairing takes place. Namely, the affinity between \ce{Li+}-\ce{NO3-}  ion pairs (CIP) in system \Romannum{4}a is large, resulting in correlated diffusion, meanwhile, \ce{Li+} and \ce{TFSI-} ions in system \Romannum{4}b are well separated by solvents and contribute to higher conductivity than in system \Romannum{4}a. Since the MD conductivities are significantly lower than the experimental values, it seems that ion pairing is a bit overestimated in our simulations. The conductivity from experimental measurements also show a decrease as \ce{LiNO3} molar concentration increases (Table \stref{ext1-table:S-sigma-expt}). This all are clear signatures of the fact that ionic conductivity is affected by not only ionic strength but also ion-specific pairing. 

We are now in the position to interpret and predict interesting properties of these state-of-the-art battery electrolytes and how they depend, for example, on the solvent composition or temperature.  Furthermore, we can also obtain a deeper microscopic insight into structural details, e.g., the detailed composition of the first solvation shell of lithium. Hence, in the following, we concentrate exemplarily on the self-diffusion coefficient of the molecules and their temperature dependence and discuss structural features of the first \ce{Li+} solvation shell in system IVa. 
\begin{center}
 \begin{table}[h!]
\caption{Fitting parameters $\ln D_0$ and $\Delta E_\trm{a}$ according to the Arrhenius law for the self-diffusion coefficients in system \Romannum{4}a.}
 \centering
 \begin{tabular}
 {@{}| K{1.5cm} | K{2.5cm}  | K{2.5cm}  |@{}} 
\hline\Tstrut
 Molecules 	& {$\ln D_0$} 	& {$\Delta E_\trm{a}$ (\si{\kJ\per\mol})} \\
 \hline\Tstrut
\ce{Li+} 	& 6.3(3)  & 12.1(1) \\
\ce{NO3-} 	& 6.0(3)  & 11.3(1) \\
\ce{TFSI-} 	& 7.4(3)  & 14.5(2) \\
DME 		& 7.5(3)  & 12.8(1) \\
DOL 		& 7.7(3)  & 12.8(1) \\
\hline
\end{tabular}
\label{table:sys4-d-arrhenius-fit}
\end{table}
\end{center}

\begin{figure}[h!]
\begin{center} 
\includegraphics[width=0.8\linewidth]{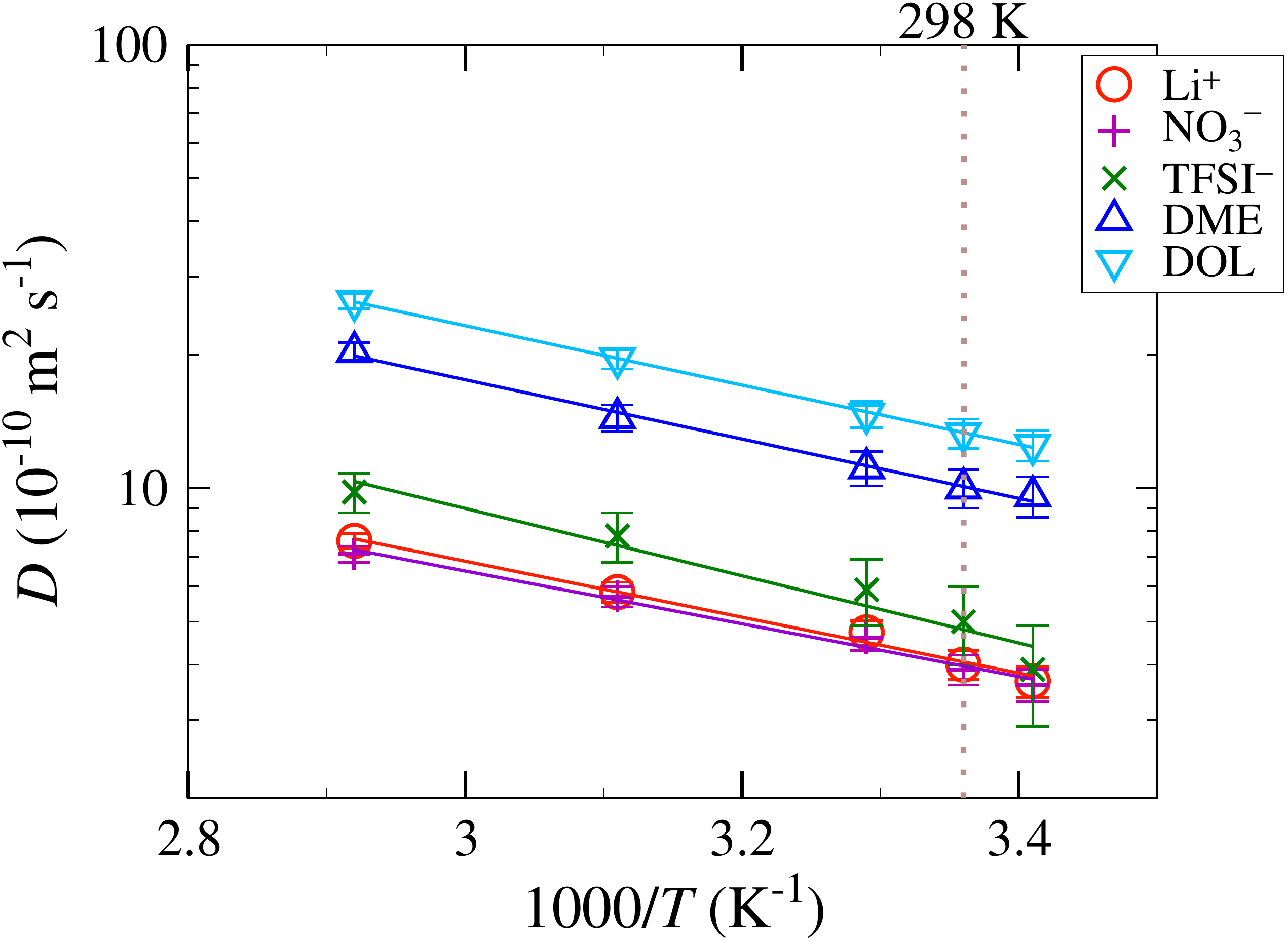}
\caption{Diffusion coefficients of ions and solvent molecules in system \Romannum{4}a as a function of inverse temperature in a log-lin representation. 
 They all obey the Arrhenius behavior given by eq.~(\ref{eq:Arr}). The vertical dotted line indicates $T=298$~K.}
\label{fig:d-5}
\end{center}
\end{figure}
Table~\ref{table:sys34} (bottom) also summarizes the results for the individual self-diffusion coefficients in system IVa. 
Compared with the dilute electrolyte systems II and III, the diffusion coefficients of both ions and solvent are substantially lower. Overall, this is a consequence of the higher viscosity of system IVa and ion pairing. The temperature-dependence of ionic diffusion is presented in Fig.~\ref{fig:d-5}, which shows an Arrhenius plot for the diffusion coefficients according to the standard law
\begin{equation}\label{eq:Arr}
 D(T) = D_0 \exp \bigg( - \frac{ \Delta E_\trm{a}}{k_\mathrm{B} T} \bigg),
\end{equation}
where $\Delta E_\trm{a}$ is the activation energy for diffusion. The individual fitting parameters $\ln{D_0}$ and $\Delta E_\trm{a}$ are summarized in Table~\ref{table:sys4-d-arrhenius-fit}.  As can be seen, the diffusion (i.e., ionic mobilities in these experimentally relevant systems) can increase almost by a factor 2 or 3 when going from room temperature to relatively hot operating temperatures close to the solvent boiling temperatures (358~K for DME and 347~K for DOL). 

\begin{figure}[ht!]
\begin{center} 
\includegraphics[width=0.7\linewidth]{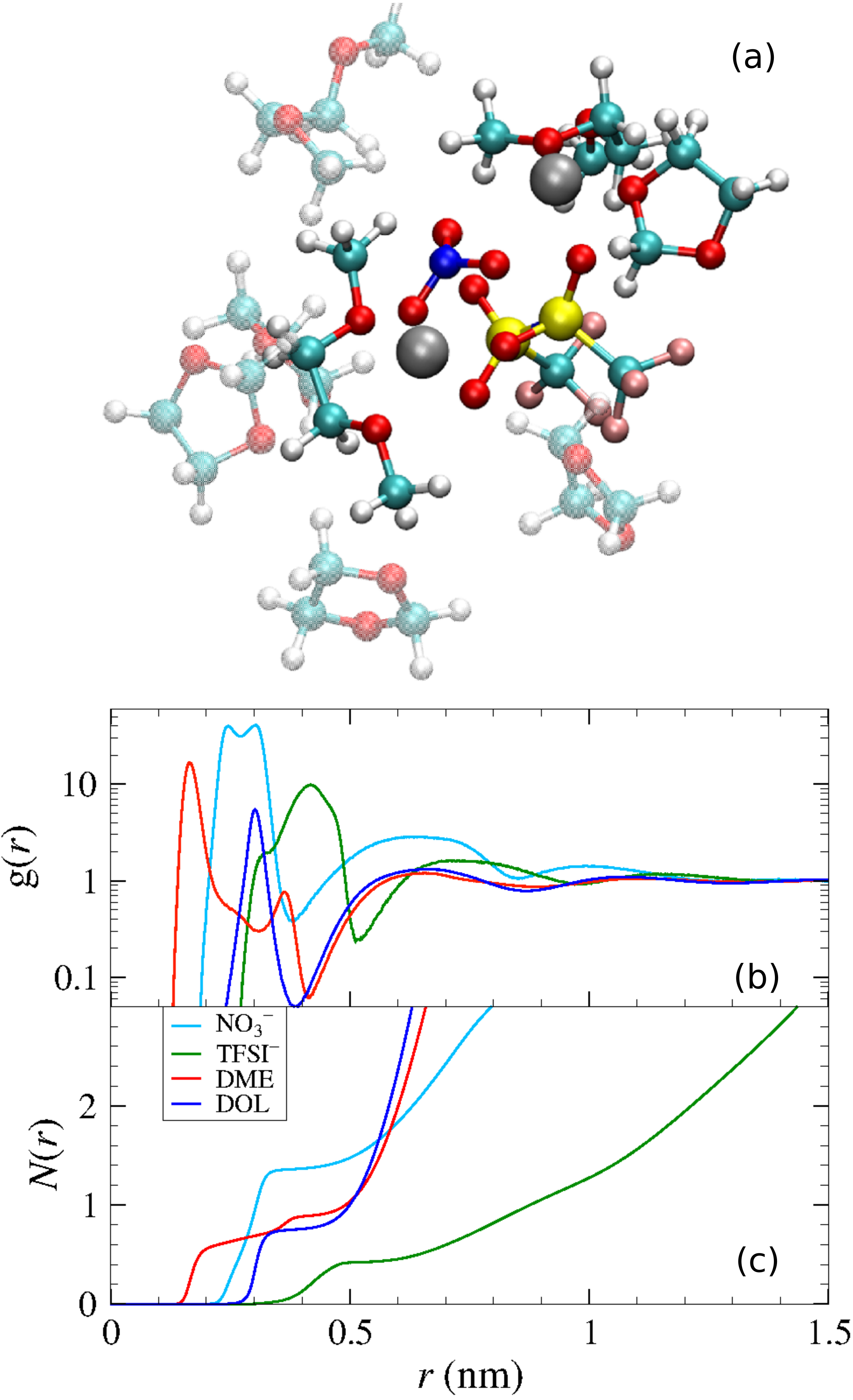}
\caption{(a) Simulation snapshot of molecules surrounding \ce{Li+} ions (gray spheres) in their first solvation shell in system \Romannum{4}a. (b) Center-of-mass RDF and (c)~the coordination number $N(r)$ of ions and solvent molecules around a single \ce{Li+} ion as a function of distance $r$ in system IVa. Note the log-lin presentation in panel (b).}
\label{fig:sys4-log-n-all}
\end{center}
\end{figure}

We finally turn to the structural description of system IVa. A snapshot of a representative configuration in the first solvation shell of Li$^+$ is shown in Fig.~\ref{fig:sys4-log-n-all}(a). The RDFs between the center-of-mass of the individual anions and solvent molecules around a \ce{Li+} ion are presented in Fig.~\ref{fig:sys4-log-n-all}(b) in a log-lin representation. A distinct solvent composition and layering within the first solvation shell is exhibited. At closest distance is the DME solvent at about 0.165~nm, followed by a large nitrate peak at about 0.25--0.3~nm and the DOL at about \SI{0.3}{\nm}. The large TFSI$^-$ has its center-of-mass a bit more outwards, peaking at about \SI{0.42}{\nm}. The first solvation shell according to these distributions has a radius of about \SI{0.5}{\nm} (see also Figs~\sfref{ext1-fig:rdf-4-2}). This running coordination number of the molecules around the \ce{Li+} ion is displayed in\ panel~(c) of the same figure. The coordination numbers in the first solvation  shell (i.e., within $\simeq$ \SI{0.5}{\nm})  are about 1.1 for DME, 0.7 for DOL, 1.3 for \ce{NO3-}, and 0.4 for \ce{TFSI-}, on average. Altogether this makes 3.2 molecules in the first solvation shell. Most qualitative structural features of system \Romannum{4}a thus resemble those of the dilute systems discussed before (see also Figs.~\sfref{ext1-fig:rdf-4-2} which, for instance, exhibit a similar structure of RDFs between system \Romannum{2}a and \Romannum{4}a).

\section{Summary and concluding remarks}
\label{sec:conclusions}

In this work, we constructed an efficient molecular model for state-of-the-art Li/S battery electrolytes and solvents that reproduces a variety of experimentally observable structural and dynamical features. We validated  it with various reference systems at hand, in particular in those limits where neat experimental data were available.  For example, the density, dielectric constant, viscosity, and diffusion coefficient of solvent mixtures DME/DOL are satisfactorily reproduced for all molar ratios.  The \ce{Li+} solvation structure and pair association with \ce{NO3-} and \ce{TFSI-} anions in DME were found consistent with experimental data. The ion mobility and conductivity in 1:20 salt--solvent systems as well agreed with experimental measurements.  Finally, the physical properties, such as \ce{Li+} solvation environment and diffusivity of the full state-of-the-art Li/S battery electrolytes were in detail investigated and gave unprecedented structural insight in the composition of the important first solvation shell of the \ce{Li+} ion. Apart from the fundamental insights provided, our model will thus serve as a basis for efficient future modelings of electrolyte structure, conductivity, capacity, etc. in various electrolyte solvent compositions in porous electrode confinements and interfaces. With that, it will provide a guidance for the development of modern Li/S batteries and related systems. 

\section*{Acknowledgments}
The authors thank Won Kyu Kim, Rafael Roa Chamorro, and R. Gregor Wei{\ss} for useful discussions.


\end{document}